\documentclass[prb,twocolumn,showpacs,superscriptaddress,citeautoscript]{revtex4}
\usepackage{graphicx}
\usepackage{amsmath,amssymb}
\begin{document}
\title{Vibrational effects in laser driven molecular wires}
\author{J\"org Lehmann}
\affiliation{Institut f\"ur Physik, Universit\"at Augsburg,
        Universit\"atsstra\ss e~1, D-86135 Augsburg, Germany}
\author{Sigmund Kohler}
\affiliation{Institut f\"ur Physik, Universit\"at Augsburg,
        Universit\"atsstra\ss e~1, D-86135 Augsburg, Germany}
\author{Volkhard May}
\affiliation{Institut f\"ur Physik, Humboldt-Universit\"at zu Berlin,
        Newtonstra\ss e~15, D-12489 Berlin, Germany}
\author{Peter H\"anggi}
\affiliation{Institut f\"ur Physik, Universit\"at Augsburg,
        Universit\"atsstra\ss e~1, D-86135 Augsburg, Germany}
\date{\today}
%
\begin{abstract}

The influence of an electron-vibrational coupling on the laser control of
electron transport through a molecular wire that is attached to several electronic leads is
investigated. These molecular vibrational modes induce an effective
electron-electron interaction. In the regime where the wire electrons couple weakly to both
the external leads and the vibrational modes, we derive within a Hartree-Fock
approximation a nonlinear set of quantum kinetic equations.
The quantum kinetic theory is then used to evaluate the laser driven, time-averaged
electron current through the wire-leads contacts.  This novel formalism is applied to two
archetypical situations in the presence of electron-vibrational effects, namely,
(i) the generation of a ratchet or pump current
in a symmetrical molecule by a harmonic mixing field and (ii) the laser
switching of the current through the molecule.

\end{abstract}
\maketitle

\renewcommand{\d}{\mathrm{d}}
\newcommand{\e}{\mathrm{e}}
\newcommand{\Tr}{\mathop\mathrm{Tr}}
\newcommand{\drangle}{\rangle\!\rangle}
\newcommand{\dlangle}{\langle\!\langle}

\def\ldef{=}
\def\rdef{=}

\newcommand{\T}{\mathcal{T}}
\newcommand{\wire}{\mathrm{wire}}
\newcommand{\leads}{\mathrm{leads}}
\newcommand{\wl}{\text{wire-leads}}
\newcommand{\ev}{\text{el-vib}}
\newcommand{\ph}{\mathrm{vib}}
\newcommand{\Hcontacts}{H_\mathrm{contacts}}
\newcommand{\Henv}{H_\mathrm{env}}
\newcommand{\Hcoupl}{H_\mathrm{coupl}}
\newcommand{\deltaeps}[3]{{\Delta_{#1#2,#3}}}
\newcommand{\erho}{P}

{\renewcommand{\i}{\mathrm{i}}
%
\section{Introduction}
%
In recent years, considerable experimental progress in the
determination the current-voltage characteristics of molecular wires
has been achieved \cite{Hanggi2002a,nitzan03,Seideman2003a,Heath2003a}.  In these experiments,
a single molecule is contacted with two nanoelectrodes such that a
transport voltage can be applied.  One recent measurement, for
example, focused on the influence of the chemical anchor group which
couples the molecule, an oligothiophene derivative, to the
electrode~\cite{thiophene03}. In another experiment, the
length-dependence of the current through a DNA strand \cite{dna03} was
investigated.  Most descriptions of such transport experiments rely on
generalizations of the scattering approach put forward by
Landauer~\cite{Landauer1957a,Datta1995a}.  Presently, the main theoretical focus
lies on the \textit{ab-initio} computation of the orbitals relevant
for the motion of excess charges through the molecular
wire~\cite{DiVentra2000a, DiVentra2002a, Xue2002a, Damle2002a,
  Heurich2002a}.

Another line of research employs rather generic models to gain a
qualitative understanding of the transport mechanisms involved.  An
important problem addressed in this way is the conduction mechanism
in the presence of electron-vibrational coupling~\cite{ratner98, bishop99,
Emberly2000a, shi00, Segal2000a, Ness2001a, schoeller01, Petrov2001a, Petrov2002a,
May2002a, Petrov2004a}.  With increasing strength of the coupling between the
wire electrons and the vibrations, the electrons tend to localize on single
wire units. Correspondingly the transport mechanism changes from a purely
coherent transfer to a sequential hopping process.  The related transfer
rates are known from the theory of non-adiabatic electron transfer
\cite{maykuehn04}. Such a special regime of charge transmission can also be
described within a Redfield theory after carrying out a
so-called polaron transformation \cite{Segal2002a, Petrov2004a}. If the wire
has to be described by spatially extended molecular orbitals, sequential
transfer proceeds by jumps from one electrode into a wire orbital,
possibly followed by an intra-wire relaxation, and then by a jump from a
wire orbital into the other electrode.  In contrast, the so-called
superexchange mechanism enables a direct transfer from one electrode to the
other.  The crossover from superexchange to a sequential transfer mechanism has been studied in
Refs.~~\onlinecite{Segal2000a, Segal2001a, Petrov1995a, Petrov2001a, May2002a,
Petrov2004a}.  In that context, a treatment of inelastic scattering
processes profits from an unified theory of the electron transfer through
molecular donor-acceptor complexes~\cite{Petrov2001b}.

Molecular wires illuminated by laser fields have been proposed for the
investigation of ac transport effects like coherent quantum
ratchets~\cite{Lehmann2002b, Lehmann2003b}, optical control of current
and noise~\cite{Lehmann2003a, Camalet2003a}, and resonant current
amplification~\cite{Keller2002a,Kohler2002a}.  The appropriate
treatment of these ac phenomena is based on Floquet theory, which
allows to take into account the action of the time-dependent field
exactly.  A Floquet scattering approach for the fully coherent
transport regime has been developed~\cite{Camalet2003a}, but it cannot
be generalized straightforwardly to the case with additional
electron-vibrational coupling. Better suited for this situation is a
quantum kinetic equation formalism which, however, is perturbative in both the
wire-lead coupling and the electron-vibrational coupling.

In the absence of an external transport voltage, a driving field can
induce a so-called pump or ratchet
current~\cite{Switkes1999a,Reimann2002a,Reimann2002b,Astumian2002a}.  The same happens even in
perfectly symmetric conductors if one adds a higher harmonic to the
driving field.  The investigation of the corresponding effect for the
motion of a particle in a tight-binding lattice revealed that the
resulting current depends sensitively on the relative phase between
the two components of the driving.  For this system, two limits have
been studied: the fully coherent dynamics and the overdamped Brownian
motion.  The dependence of the current on the relative phase is in
each case qualitatively different \cite{Goychuk1998b,Goychuk2001a}.  The present
model has the advantage that it enables the study of such effects also in
the crossover regime between the purely coherent and the purely
incoherent transport.

A further intriguing phenomenon in driven transport is the suppression of
the dc current caused by properly tailored ac fields \cite{Lehmann2003a}.
This effect is the transport counterpart of the so-called coherent
destruction of tunneling (CDT) found in bistable potentials without any
connection to external leads \cite{Grossmann1991a,Grossmann1991b,Bavli1992a}.  When coupling the
bistable system to a heat bath, tunneling becomes a transient, see
Ref.~~\onlinecite{Grifoni1998a} for a review.  The same is true for coherent
destruction of tunneling: ultimately, the driving-induced localization
decays via dissipative transitions \cite{Grifoni1998a}.  Here, we address
the role of dissipation for the corresponding transport effect.

The paper is organized as follows: In Sect.~\ref{sec:model}, we introduce
a model for the laser driven molecule coupled to several leads and
vibrational degrees of freedom.  Subsequently, in
Sect.~\ref{sec:kinetic_equation}, we derive a kinetic equation approach in
the Floquet basis.  An expression for the resulting current through the
molecule is derived in Sect.~\ref{sec:current}. Finally, in
Sect.~\ref{sec:applications}, the formalism is applied to study the
influence of the vibrational coupling on non-adiabatic pumping and coherent
current suppression.
%
%
\section{The model}
\label{sec:model}
%
In the following, we consider a molecular wire which is attached to a number of
electrodes and which is driven by an
externally applied ac field. We neglect hole transport and start with
the Hamiltonian
\begin{equation}
H_0(R) = E_0 + H_{\ph}\,,
\end{equation}
of the neutral wire, i.e., in the absence of excess electrons injected
via the electrodes.  Here, $R$ denotes the set of all involved
vibrational coordinates with equilibrium configuration $R_0$ and $E_0
= H_0(R_0)$ is the electronic ground-state energy of the neutral wire
(set equal to zero in the following). In a representation by normal
modes with mode index $\xi$, we find the vibrational Hamiltonian
\begin{equation}
  \label{3.2kb}
 H_{\ph} =
\sum_{\xi}  \hbar \omega_{\xi} \, b^\dagger_{\xi} b_{\xi}
\end{equation}
with the usual harmonic oscillator operators $b^\dagger_{\xi}$ and
$b_{\xi}$.  The normal modes may be delocalized over the whole wire or
may be localized on specific wire units. We assume that they always
remain in thermal equilibrium and are, thus, characterized by the Bose
distribution $n_\mathrm{B}(\hbar\omega_{\xi})$.

We describe the presence of an excess electron on the wire in the
representation of $N$ localized wire orbitals $|n\rangle$,
$n=1,\dots,N$, and the corresponding Hamiltonian $H_{nn'}(R, t)$.  In
the absence of the external driving, the eigenstates of $H_{nn'}$ are
the so-called LUMOs (lowest unoccupied molecular orbitals).  To
account for electron-vibrational coupling this quantity is expanded
with respect to deviations from the vibrational equilibrium
configuration $R_0$. To lowest order this results in
\begin{equation}
  \label{3.2}
H_{n n'}(R, t) = H_{n n'}(t)  +  \sum_{\xi} K_{n n', \xi} \,
( b_{\xi} + b^\dagger_{\xi}) \,
\end{equation}
The matrix $H_{n n'}(t) = H_{n n'}(R_0, t)$ taken at the vibrational
equilibrium configuration $R_0$ includes the interaction with the
external field. The $\T$-periodic time-dependence $H_{nn'}(t) =
H_{nn'}(t+\T)$ models the action of the external field on the excess
electrons when moving through the wire. We have in mind a dipole-type
coupling between electron and field, i.e., a contribution $e\,
\mathbf{E}(t) \mathbf{x}_n\, \delta_{n, n'} $ to the Hamiltonian
$H_{nn'}(t)$, where $\mathbf{E}(t)$ is the electric field strength,
$\mathbf{x}_n$ the position of site $n$, and $-e$ the electron charge.
The considered action of an external field may in principle also
induce a heating of the electron gas in the metallic leads as well as
a thermal expansion of the leads.  Of course, this can be of
importance for any experimental realization of the studied mechanism.
Nevertheless, the present approach will not attempt to account for
such effects.

We assume that the whole set of vibrational coordinates discerns into
subsets, labelled by an index $\nu$.  Therefore, we write the
electron-vibrational coupling in Eq.~\eqref{3.2} as
\begin{equation}
\sum_{\xi} K_{n n', \xi} \,
( b_{\xi} + b^\dagger_{\xi}) = \sum_\nu X_{n n' \nu} \sum_{\xi\in\nu}  M_{\nu  \xi } \, ( b_{\xi} + b^\dagger_{\xi})\,,
\end{equation}
where the notation $\xi\in\nu$ indicates that the summation runs only
over the modes in the corresponding subset.  The related spectral
densities read
\begin{equation}
  \label{3.2l-1}
  D_{\nu}(\omega) = \frac{\pi}{\hbar} \sum_{\xi \in \nu} | M_{\nu \xi} |^2\,
  \delta(\omega - \omega_{\xi}) \,.
\end{equation}
For notational convenience, we define the spectral density for $\omega < 0$
by $D_{\nu}(-\omega) = - D_{\nu}(\omega)$.
In our numerical calculations, we restrict ourselves to a situation
where each orbital $|n\rangle$ couples to exactly one of these
subsets, i.e.,
\begin{equation}
  \label{3.2l}
  X_{nn'\nu} = \delta_{nn'} \, \delta_{n \nu} \,,
\end{equation}
and assume identical Ohmic spectral densities $D_\nu(\omega) = \kappa \hbar \omega$,
where the dimensionless coupling strength~$\kappa$ is identical for all
sites. This model has been employed recently for the description of
dephasing and relaxation in (time-independent) bridged molecular
wires \cite{Segal2000a, Segal2001a}.

In order to take the exclusion principle properly into account, we
employ a many-electron description based on a second quantized
notation.  Neglecting the Coulomb interaction among the electrons, the
wire electrons are described by the Hamiltonian
\begin{equation}
  \label{3.2a}
  H_{\wire}(t) = \sum_{n,n'}  H_{nn'}(t) \, c^\dagger_n c_{n'} \,.
\end{equation}
The operators $c_n$ and $c^{\dagger}_n$ annihilate and create,
respectively, an electron in the orbital $|n\rangle$,
$n=1,\dots,N$.  We presume vanishing overlap among different orbitals such
that the annihilation and creation operators obey the standard
anti-commutation relations $[c_n,c^\dagger_{n'}]_+=\delta_{nn'}$.
The electron-vibrational coupling Hamiltonian assumes the form
\begin{equation}
  H_\ev = 
  \sum_\nu X_{n n' \nu} \sum_{\xi\in\nu}  M_{\nu  \xi }
  \, (b_\xi + b^\dagger_\xi)\,.
\end{equation}

Next, we consider the coupling of the wire to $L$ different macroscopic
electronic leads described by the Hamiltonian
\begin{equation}
  \label{3.2c}
  H_\leads = \sum_{{k}, \ell} \epsilon_{\mathbf{k}\ell} \,
  c^\dagger_{\mathbf{k} \ell} \, c_{\mathbf{k} \ell}\,.
\end{equation}
The electrons in each lead~$\ell$ are labelled by wavevectors
$\mathbf{k}$ referring to bulk or surface states with energy
$\epsilon_{\mathbf{k} \ell}$. All lead states are mutually orthogonal
(and also orthogonal to the wire states) and, therefore, the creation
and annihilation operators $c_{\mathbf{k} \ell}$ and
$c^\dagger_{\mathbf{k} \ell}$, respectively, obey the standard
anti-commutation relations.  We assume that the lead electrons stay in
thermal equilibrium and are, thus, described by the Fermi
distributions $f(\epsilon_{\mathbf{k} \ell}- \mu_\ell)$ with a common
temperature~$T$ but possibly different electro-chemical potentials
$\mu_\ell$.

The coupling of each lead to exactly one of the suitably labelled
molecular orbitals is described by the Hamiltonian
\begin{equation}
  \label{3.2e}
  H_\wl = \sum_{{\bf k}, \ell} V_{{\bf k} \ell} \, c^\dagger_{{\bf k} \ell} \, c_\ell + \text{h.c.}
\end{equation}
with $V_{{\bf k} \ell}$ being the tunneling matrix elements.  As it
will turn out, the coupling to the leads is completely characterized
by its spectral spectral density
\begin{equation}
  \label{3.2f}
  \Gamma_\ell(\epsilon) \ldef  \frac{2\pi}{\hbar} \sum_{\bf k} |V_{{\bf k} \ell}|^2 \,
  \delta(\epsilon-\epsilon_{{\bf k} \ell})\ ,
\end{equation}
which becomes a continuous function of the energy $\epsilon$ if the
lead states are dense. Since we
are mainly interested in the behavior of the molecular wire itself and not
in the details of the lead-wire coupling~\cite{Yaliraki1998a}, we will assume
for all numerical calculations that the conduction band width
of the leads is much larger than all remaining relevant energy scales.
In this so-called wide-band limit, the spectral densities are constants,
$\Gamma_\ell(\epsilon) = \Gamma_\ell$.

The dynamics of the present model is now fully specified by the Hamiltonian
\begin{equation}
  \label{3.1}
  H(t) = H_\wire(t) + H_\ph + H_\leads + H_\ev + H_\wl\,.
\end{equation}
In the following, we start from the uncoupled subsystems $H_\wire(t)+
H_\ph + H_\leads$ and treat the influence of both the wire-lead
coupling and the electron-vibrational coupling, subsumed in the Hamiltonian
\begin{equation}
  \label{3.2g}
  \Hcoupl = H_\ev + H_\wl\,,
\end{equation}
within a master equation approach in second order perturbation theory.
%
\section{Quantum kinetic equation approach}
\label{sec:kinetic_equation}
%
Within this work, we focus on the case of weak and intermediately
strong coupling of the electrons on the wire to both the vibrational
environment and the electronic states in the leads. Thus, it is
favorable to choose a description in terms of the reduced density
operator $\varrho_\wire(t)$ of the wire electrons which follows from the
total density operator by tracing out those degrees of freedom which
correspond to vibrations and lead electrons.  Then, the derivation of
a closed equation for $\varrho_\wire(t)$ to second order in $\Hcoupl$
represents a standard procedure of dissipative quantum dynamics (for a
review, see, e.g., Ref.~~\onlinecite{maykuehn04}). We neglect initial
correlations between the wire electrons and the environmental degrees
of freedom, which stay in thermal equilibrium, and do not account for
non-Markovian dissipative effects. The resulting quantum master
equation thus reads
\begin{multline}
\label{3.8}
\dot\varrho_\wire(t) =
 -\frac{\i}{\hbar}[H_{\rm wire}(t),\varrho_\wire(t)] \\
 -\frac{1}{\hbar^2}\int\limits_0^{\infty} \!\!\d\tau
  \Tr\nolimits_\mathrm{env} [\Hcoupl,
  [\widetilde{H}_\mathrm{coupl}(t-\tau,t),
  \varrho_\wire(t)\otimes\varrho_\mathrm{env,eq}]]  .
\end{multline}
The trace refers to all environmental states, i.e., the electronic
states of the leads as well as the wire vibrations, and the operator
$\widetilde{H}_\mathrm{coupl}(t,t') \ldef U^\dagger_0(t,t') \Hcoupl
U_0(t,t')$ describes the coupling Hamiltonian~\eqref{3.2g} in the
interaction representation.  The related zeroth-order time evolution
operator
\begin{multline}
\label{3.4}
U_0(t,t') \\
= {\textstyle T} \exp\left(-\frac{\i}{\hbar}\int_{t'}^t \d t''\,
[H_\wire(t'')+H_\ph+H_\leads]\right)
\end{multline}
is responsible for the dynamics of the uncoupled subsystems.  In this way the
external driving field not only determines the coherent dynamics of
the wire electrons but also the dissipative part of the master
equation~\eqref{3.8}~\cite{Kohler1997a,Kohler1998a}.

Note that Eq.~\eqref{3.8} still determines the dynamics of the full
many-particle density matrix of the wire electrons. Later on, we will
derive an equation of motion for the single wire-electron density
matrix defined as
\begin{equation}
  \label{el-density-matrix}
\erho_{n n'}(t) =
\Tr\nolimits_\mathrm{el} \,[ \varrho_\wire(t) \, c^\dagger_{n'}\, c_n ] \ ,
\end{equation}
where the trace runs over the many-particle states of the wire
electrons. It will be demonstrated below that in the presence of an
electron-vibrational coupling, a closed equation for $\erho_{n
  n'}(t)$ can only be obtained when an approximation is carried out
for the two-electron density matrices by employing a  decoupling scheme.
We will detail this point when introducing the Hartree-Fock
approximation in Section~\ref{Floquet_representation_of_the_electron_vibrational_coupling}.
However, before doing so, we will briefly review the Floquet method
for the treatment of the explicit time-dependence appearing in the
propagator~\eqref{3.4}.

%
\subsection{Introduction of Floquet states}
\label{Incoherent_dynamics}
%
For the evaluation of Eq.~\eqref{3.8} it is essential to use an exact
expression for the zeroth-order time evolution operator $U_0(t,t')$.
The use of any approximation bears the danger of generating artifacts,
which, for instance, may lead to a violation of fundamental
equilibrium properties~\cite{maykuehn04, Novotny2002a}. In the present
case, the only non-trivial contribution to the propagator~\eqref{3.4}
stems from the periodically time-dependent wire Hamiltonian
$H_{nn'}(t) = H_{nn'}(t+\T)$. A proper tool for the
efficient computation of the corresponding propagator is Floquet
theory~\cite{Shirley1965a, Sambe1973a, Fainshtein1978a, Grifoni1998a},
which is based on the fact that there exists a complete set of
solutions of the form
\begin{equation}
  \label{2.2}
  |\Psi_\alpha(t)\rangle =
  \e^{-\i\epsilon_\alpha t/\hbar} \,|\Phi_\alpha(t)\rangle\ , \quad
  |\Phi_\alpha(t)\rangle= |\Phi_\alpha(t+\T)\rangle
\end{equation}
with the so-called quasienergies $\epsilon_\alpha$ and corresponding
Floquet modes $|\Phi_\alpha(t) \rangle$. The Floquet modes fulfill the
eigenvalue equation
\begin{equation}
  \label{3.11a}
  \Big(\sum_{n,n'}|n\rangle H_{nn'}(t) \langle n'|-\i\hbar\frac{\d}{\d t}\Big)
  |\Phi_{\alpha}(t)\rangle = \epsilon_\alpha |\Phi_{\alpha}(t)\rangle\ .
\end{equation}
The practical usefulness of the Floquet ansatz~\eqref{2.2} is rooted in the
fact that the Floquet modes are periodic functions of time $t$ and can
therefore be decomposed in a Fourier series
\begin{equation}
  \label{2.8b}
  \begin{split}
    |\Phi_\alpha(t)\rangle & = \sum_k \e^{-\i k\Omega t} \,|\Phi_{\alpha,k}\rangle\,,\\
    |\Phi_{\alpha,k}\rangle  & \ldef  \frac{1}{\T} \int_0^\T \!\d t\, \e^{\i k \Omega t} \,
    |\Phi_\alpha(t)\rangle\,.
  \end{split}
\end{equation}
This representation makes explicit that each
quasi\-energy~$\epsilon_\alpha$ is equivalent to the quasienergies
\begin{equation}
  \label{2.11a}
  \epsilon_{\alpha,k}\ldef \epsilon_\alpha + k\hbar\Omega\,,
\end{equation}
where $k$ is an arbitrary integer.  Thus, we can restrict ourselves to
states with eigenvalues in one Brillouin zone, $E-\hbar\Omega/2 \le
\epsilon_\alpha < E+\hbar\Omega/2$.

It is now convenient to define a ``Floquet picture''
via the time-dependent transformation of the fermionic creation and
annihilation operators
\begin{equation}
\label{3.12}
c_\alpha(t) = \sum_n \langle\Phi_\alpha(t)|n\rangle\, c_n\ .
\end{equation}
The inverse transformation
\begin{equation}
c_n = \sum_{\alpha} \langle n|\Phi_\alpha(t)\rangle\,c_\alpha(t)
\label{3.13}
\end{equation}
follows from the mutual orthogonality and the completeness of the
Floquet states at equal times~\cite{Grifoni1998a}.  Note
that the right-hand side of Eq.~(\ref{3.13}) becomes $t$-independent
after the summation over $\alpha$.  In the interaction picture, the operators
$c_\alpha(t)$ obey
\begin{equation}
    \tilde c_\alpha(t,t')
    =U_0^\dagger(t,t')\,c_\alpha(t')\,U_0(t,t')
    =\e^{-\i\epsilon_\alpha (t-t')/\hbar}\, c_\alpha(t')\ .
    \label{3.13a}
\end{equation}
This is readily verified by differentiating expression~\eqref{3.13a}
with respect to $t$ and using the fact that
$|\Phi_\alpha(t)\rangle$ is a solution of the eigenvalue equation
(\ref{3.11a}). The proof is completed by noting that Eq.~\eqref{3.13a}
fulfills the initial condition $\tilde c_\alpha(t',t') = c_\alpha(t')$.

It is advantageous to change to the ``Floquet representation'' of the
single (wire) electron density operator
\begin{equation}
  \label{3.15}
  \begin{split}
  \erho_{\alpha \beta}(t) &= \sum_{n, n'} \langle n | \Phi_\alpha(t) \rangle
  \langle\Phi_\alpha(t)|n'\rangle\, \erho_{n n'}(t) \\
 &= \Tr\nolimits_\mathrm{el} \,[ \varrho_\wire(t)\,  c^\dagger_{\beta}(t)\, c_{\alpha}(t) ]
   = \langle c^\dagger_{\beta}(t) c_{\alpha}(t) \rangle_t\ .
  \end{split}
\end{equation}
After some algebra, we obtain for the dynamics of these expectation
values the expression
\begin{multline}
\label{3.8a}
  \frac{\d}{\d t} \erho_{\alpha\beta}(t)
  = {}
  -\frac{\i}{\hbar} (\epsilon_\alpha -  \epsilon_\beta)\,
  \erho_{\alpha\beta}(t)
  \\
 -\frac{1}{\hbar^2}\int\limits_0^{\infty} \!\!\d\tau\,
  \langle [[c^\dagger_{\beta}(t) c_{\alpha}(t), \Hcoupl],
  \widetilde{H}_\mathrm{coupl}(t-\tau,t)]
  \rangle_t\ .
\end{multline}
Obviously, the canonical transformation~\eqref{3.12} to the basis of
the Floquet operators $c_\alpha(t)$ has diagonalized the coherent part
of the master equation~\eqref{3.8a} and the only task left is the
evaluation of the incoherent contribution.  Here, we use the fact that the
contributions resulting from the coupling of the wire electrons to the
electronic leads and to the wire vibrations can be treated separately.
This is possible due to the assumption that the lead electrons and the
vibrations remain uncorrelated at all times. Finally, we will
obtain as a main result a quantum
kinetic equation of the form
\begin{equation}
  \label{3.15a}
  \dot\erho_{\alpha\beta} =
  -\frac{\i}{\hbar} (\epsilon_\alpha -  \epsilon_\beta)
  \erho_{\alpha\beta}
  +
  \dot\erho_{\alpha\beta}\big\vert_\wl
  +
  \dot\erho_{\alpha\beta}\big\vert_\ev
  \,.
\end{equation}
Specific expressions for the second and third term on the right-hand side
of the last equation will be derived in the following sections.

%
\subsection{Floquet-representation of the wire-leads coupling}
\label{Floquet_representation_of_the_wire_leads_coupling}
%
For the evaluation of the contribution of the wire-leads coupling to
the kinetic equation~\eqref{3.15a}, we have to evaluate the integral
in Eq.~\eqref{3.8a} for the corresponding term in the coupling
Hamiltonian~\eqref{3.2g}. Using the relation $\tilde
c_{q\ell}(t-\tau,t) = \exp(\i\epsilon_{q\ell}\tau/\hbar) \,
c_{q\ell}$, we obtain
\begin{equation}
  \label{3.16}
  \begin{split}
  &
  \dot\erho_{\alpha\beta}\big\vert_\wl
  =
  \sum_{\ell=1}^L
  \int\limits_0^\infty \frac{\d\tau}{\hbar}
  \int\!\frac{\d\epsilon}{2\pi}\,
  \Gamma_\ell(\epsilon)
  \\
  \times &{}
  \Big\{
  \e^{-\i\epsilon \tau/\hbar}
  \Big[
    \langle
      [[c^\dagger_\beta(t)\, c_\alpha(t), c^\dagger_\ell],
      \tilde{c}_\ell
      ]_+
    \rangle_t\,
    f(\epsilon-\mu_\ell)
    \\
    &
    \qquad\qquad
    -\langle
      [c^\dagger_\beta(t)\, c_\alpha(t), c^\dagger_\ell]\,
      \tilde{c}_\ell
    \rangle_t
  \Big]
  \\
  &
  -
  \e^{\i\epsilon \tau/\hbar}
  \Big[
    \langle
      [[c^\dagger_\beta(t)\, c_\alpha(t), c_\ell],
      \tilde{c}^\dagger_\ell
      ]_+
    \rangle_t\,
    f(\epsilon-\mu_\ell)\\
    &
    \qquad\qquad
    -
    \langle
      \tilde{c}^\dagger_\ell
      [c^\dagger_\beta(t)\, c_\alpha(t), c_\ell]
    \rangle_t
  \Big]
  \Big\}\ .
\end{split}
\end{equation}
For notational compactness, the time arguments of the interaction
picture operators $\tilde{c}_\ell(t-\tau,t)$ have been suppressed.

Using the transformation~\eqref{3.13}, the commutators in Eq.~\eqref{3.16} are
readily evaluated to read
\begin{align}
  \label{3.17}
  [c^\dagger_\beta(t)\,c_\alpha(t), c_\ell] & =
  -\langle \ell|\Phi_\beta(t)\rangle\, c_\alpha(t)\ ,\\
  [c^\dagger_\beta(t)\,c_\alpha(t), c^\dagger_\ell] & =
  \langle \Phi_\alpha(t)|\ell\rangle\, c^ \dagger_\beta(t)\ .
\end{align}
Moreover, the transformation~\eqref{3.13} yields in conjunction with
Eqs.~\eqref{2.8b}, \eqref{2.11a}, and~\eqref{3.13a} the spectral
decomposition of the wire operators in the interaction picture, i.e.,
\begin{equation}
  \label{3.17a}
  \tilde{c}_\ell(t-\tau,t) =
  \sum_{\alpha k}
  \e^{-\i k\Omega t}\,
  \e^{\i\epsilon_{\alpha,k}\tau/\hbar} \langle\ell|\Phi_{\alpha,k}\rangle\,
  c_{\alpha}(t)\ .
\end{equation}
With the aid of the last two equations, one may readily carry out the time and the energy
integrations in Eq.~\eqref{3.16} to obtain
\begin{equation}
  \label{3.18}
  \begin{split}
  & \dot\erho_{\alpha\beta}\big\vert_\wl  =
  \frac{1}{2}
  \sum_{\ell=1}^L
  \sum_{kk'} {}
  \e^{\i(k'-k)\Omega t}\\
  \times
  \Big\{
  &
  \Gamma_\ell(\epsilon_{\alpha,k})
  \langle\Phi_{\alpha,k'}|\ell\rangle\langle\ell|\Phi_{\beta,k}\rangle\,
  f(\epsilon_{\alpha,k}-\mu_\ell)
  \\
  & +
  \Gamma_\ell(\epsilon_{\beta,k})
  \langle\Phi_{\alpha,k'}|\ell\rangle\langle\ell|\Phi_{\beta,k}\rangle\,
   f(\epsilon_{\beta,k}-\mu_\ell)
  \\
  & -
  \sum_{\alpha'}
  \Gamma_\ell(\epsilon_{\alpha',k})
  \langle\Phi_{\alpha,k'}|\ell\rangle\langle\ell|\Phi_{\alpha',k}\rangle\,
  \erho_{\alpha'\beta}(t)
  \\
  & -
  \sum_{\beta'}
  \Gamma_\ell(\epsilon_{\beta',k'})
  \langle\Phi_{\beta',k'}|\ell\rangle\langle\ell|\Phi_{\beta,k}\rangle\,
  \erho_{\alpha\beta'}(t)
  \Big\}\ .
  \end{split}
\end{equation}
Here, principal value terms stemming from an energy renormalization
due to the coupling to the leads have been neglected.  The terms
containing Fermi functions describe resonant tunneling of electrons
from the leads onto the wire, while the reverse processes are captured
by the terms proportional to $\erho_{\alpha\beta}(t)$.
%
\subsection{Floquet-representation of the electron vibrational coupling}
\label{Floquet_representation_of_the_electron_vibrational_coupling}
%
For ease of notation, we introduce the ``position'' and the ``force''
operators
\begin{align}
  X_\nu & \ldef \sum_{n,n'} X_{nn'\nu} \, c^\dagger_n c_{n'}
  \ ,\\
  F_\nu & \ldef \sum_{\xi_\nu}  M_{\xi_\nu}
  (b_{\xi_\nu} + b^\dagger_{\xi_\nu})\ ,
\end{align}
respectively.  Then, the vibrational contribution to the dissipative
kernel in Eq.~\eqref{3.8a} becomes
\begin{equation}
  \label{3.20}
  \begin{split}
  &\dot\erho_{\alpha\beta}\big\vert_\ev =
  \\
  &-
  \frac{1}{\hbar}
  \sum_\nu
  \int\limits_0^\infty \!\d\tau\,
  S_\nu(\tau)\,
  \langle [[c^\dagger_\beta(t)\, c_\alpha(t), X_\nu], \widetilde{X}_\nu(t-\tau,t)]\rangle_t\\
  & - \frac{\i}{\hbar}\sum_\nu
  \int\limits_0^\infty \!\d\tau\,
  A_\nu(\tau)\,
  \langle [[c^\dagger_\beta(t)\, c_\alpha(t), X_\nu], \widetilde{X}_\nu(t-\tau,t)]_+\rangle_t
  \ .
  \end{split}
\end{equation}
The symmetrized and anti-symmetrized autocorrelation functions
\begin{align}
  \label{3.21}
  S_\nu(\tau) & \ldef  \frac{1}{2\hbar} \, \langle [\widetilde{F}_\nu(\tau), F_\nu]_+
  \rangle_\mathrm{eq}
  \\
  \nonumber
  &
  =
  \int_0^\infty
  \frac{\d\omega}{\pi}\,
  D_\nu(\omega)  \coth(\hbar\omega/2k_\mathrm{B} T) \cos(\omega\tau)
  \,,
  \\
  \label{3.22}
  A_\nu(\tau) & \ldef  \frac{1}{2\i\hbar} \, \langle [\widetilde{F}_\nu(\tau), F_\nu]
  \rangle_\mathrm{eq}
  \\
  \nonumber
  &
  =
  -
  \int_0^\infty
  \frac{\d\omega}{\pi}\,
  D_\nu(\omega) \sin(\omega\tau)
  \,,
\end{align}
of the ``force'' operators fully characterize the fluctuation
properties of the wire vibrations. Note that $\coth(x) = 1/x +
\mathcal{O}(x)$ such that for an Ohmic spectral density no infrared
divergence occurs in the integral~\eqref{3.21}.

For the further evaluation of Eq.~\eqref{3.20}, we express the operator $X_\nu$ and
its interaction picture version $\tilde X_\nu(t-\tau,t)$ in terms of the Floquet picture
operators $c_\alpha(t)$ at time~$t$, obtaining
\begin{align}
  \label{3.25}
  X_\nu = {}&\sum_{\alpha,\beta}
  \sum_k
  \e^{\i k\Omega t} \,
  \bar X^\nu_{\alpha\beta,k}\,
  c^\dagger_\alpha(t)
  c_\beta(t)\,,\\
  \widetilde X_\nu(t-\tau,t) = {}&
  \sum_{\alpha,\beta}
  \sum_k
  \e^{\i k\Omega t} \,
  \e^{\i (\epsilon_\beta-\epsilon_\alpha - k\hbar\Omega)\tau/\hbar}\,
  \bar X^\nu_{\alpha\beta,k}\\
  \nonumber
  & \times  c^\dagger_\alpha(t)  c_\beta(t)\,.
\end{align}
The time-averaged coupling matrix elements in the Floquet basis have been
abbreviated as
\begin{equation}
  \label{3.24}
  \bar X^\nu_{\alpha\beta,k}\ldef
  \sum_{n,n'}
  \sum_{k'}
  \langle \Phi_{\alpha,k+k'}|n\rangle
  X_{nn'\nu}
  \langle n'|\Phi_{\beta,k'}\rangle\ .
\end{equation}
When evaluating Eq.~\eqref{3.20}, it turns out that in addition to
terms containing the single-electron density matrix
$\erho_{\alpha\beta}(t)$, two-electron expectation values of the form
$\langle c^\dagger_\delta(t)\, c^\dagger_\gamma(t)\, c_\beta(t)\,
c_\alpha(t)\rangle_t$ appear. By iteration, one thus generates a
hierarchy of equations up to $N$-electron expectation values.  To
obtain a description in terms of only the single-electron expectation
values, we employ the Hartree-Fock decoupling scheme defined by the
approximation
\begin{equation}
  \label{3.23}
  \begin{split}
  \langle c^\dagger_\delta(t)\, c^\dagger_\gamma(t)\, c_\beta(t)\, c_\alpha(t)\rangle_t
  \approx {} &
  \erho_{\alpha\delta}(t) \erho_{\beta\gamma}(t)-
  \erho_{\beta\delta}(t) \erho_{\alpha\gamma}(t)\ .
  \end{split}
\end{equation}
Clearly, such a mean-field approximation only covers certain aspects
of the full many-particle problem. Nevertheless, it offers a
description which is consistent with the second law of thermodynamics,
as we will detail in Sect.~\ref{sec:3:thermaleq}. We remark that in
principle one could also include electron-electron interaction in the
framework of the mean-field approximation~\eqref{3.23}, similar to the
approach put forward in Refs.~~\onlinecite{May1990a,May1991a}.  However,
focusing on the vibration-mediated interaction effects, we here
refrain from doing so.

For the description of the transport problem, a scattering approach,
i.e., a strict one-particle picture, is frequently
employed~\cite{Bonca1995a,Ness1999a,Emberly2000a,May2002a,Segal2000a,Segal2001a}.
Then, non-linear terms of the type~\eqref{3.23} do not appear.  The
same happens if one considers a closed system with a single electron;
then, the corresponding equation is also of the form~\eqref{3.15a} but
without the terms quadratic in
$\erho_{\alpha\beta}(t)$~\cite{Dittrich1993a,
  Kohler1997a, Fonseca2004a}.

Upon insertion of the Hartree-Fock approximation~\eqref{3.23} into
Eq.~\eqref{3.20} and disregarding again the principal value integrals,
which correspond to an energy renormalization due to the
electron-vibrational coupling, we obtain after a straightforward
calculation
\begin{widetext}
\begin{equation}
  \begin{split}
    \dot\erho_{\alpha\beta}\big\vert_\ev\!\!
    = {}&
    \frac{1}{2}
    \sum_{\alpha'\beta'}
    \Big[
    \Gamma_{\alpha\alpha'\beta\beta'}(t)
    +
    \Gamma^\ast_{\beta\beta'\alpha\alpha'}(t)
    \\
    & \phantom{\frac{1}{2}\sum_{\alpha'\beta'}\Big[}
    -\sum_{\alpha''}
    \Big(
    \Gamma^\ast_{\alpha''\beta'\alpha\alpha'}(t)
    +
    \Gamma_{\beta'\alpha'\alpha''\alpha}(t)
    \Big)
    \erho_{\alpha''\beta}
    -\sum_{\beta''}
    \Big(
    \Gamma^\ast_{\alpha'\beta'\beta''\beta}(t)+
    \Gamma_{\beta''\alpha'\beta\beta'}(t)
    \Big)
    \erho_{\alpha\beta''}
    \Big]
    \erho_{\alpha'\beta'}
    \\
    &-\frac{1}{2}\sum_{\alpha'}
    \Big[
    \sum_{\beta''}
    \Gamma_{\beta''\alpha'\beta''\alpha}(t)
    -\sum_{\alpha''\beta''}
    \Big(
    \Gamma^\ast_{\alpha''\beta''\alpha\alpha'}(t)+
    \Gamma_{\beta''\alpha'\alpha''\alpha}(t)
    \Big)
    \erho_{\alpha''\beta''}
    \Big]
    \erho_{\alpha'\beta}
    \\
    &
    -\frac{1}{2}\sum_{\beta'}
    \Big[
    \sum_{\alpha''}
    \Gamma^\ast_{\alpha''\beta'\alpha''\beta}(t)
    -\sum_{\alpha''\beta''}
    \Big(
    \Gamma^\ast_{\alpha''\beta'\beta''\beta}(t)+
    \Gamma_{\beta''\alpha''\beta\beta'}(t)
    \Big)
    \erho_{\alpha''\beta''}
    \Big]
    \erho_{\alpha\beta'}
    \ .
  \end{split}
\end{equation}
\end{widetext}
Here, we have introduced the time-dependent, complex-valued coefficients
\begin{equation}
  \begin{split}
  \Gamma_{\alpha\beta\alpha'\beta'}(t)
  \ldef
  2
  \sum_\nu
  \sum_{kk'}{}&
  \e^{\i(k'-k)\Omega t}\,
  \bar X^\nu_{\alpha\beta,k'}\,
  \bar X^\nu_{\beta'\alpha',-k}
  \\
  &
  \times
  N_\nu(\epsilon_{\alpha}-\epsilon_{\beta} + k'\hbar\Omega)
  \ ,
  \end{split}
\end{equation}
where the functions $N_\nu(\epsilon)$ are defined for each vibrational
subsystem $\nu$ by $N_\nu(\epsilon) \ldef $ $D_\nu(\epsilon/\hbar) \,
n_\mathrm{B}(\epsilon)/\hbar$.  For an Ohmic spectral density,
$D_\nu(\epsilon/\hbar) \propto \epsilon$, $N_\nu(\epsilon)$ is
well-defined in the limit $\epsilon\to0$.
%
\subsection{Rotating-wave approximation}
\label{sec:3:rwa}
%
For a very weak coupling of the wire electrons to the environmental
degrees of freedom, the coherent time-evolution dominates the
dynamics~\eqref{3.15a} of the density matrix $\erho_{\alpha\beta}(t)$.
More precisely, the largest time-scale of the coherent evolution,
which is given by the smallest quasienergy difference, and the
dissipative time-scale, which is of the order of the coupling rates
$\Gamma_\ell(\epsilon)$, are well-separated, i.e.,
\begin{equation}
  \label{3.34}
  \hbar\Gamma_\ell(\epsilon)\ll
  |\epsilon_{\alpha} - \epsilon_{\beta} + k\hbar\Omega|
  \,,\quad
  \kappa\ll1
\end{equation}
for all $\ell$, $k$, $\epsilon$ and $\alpha\ne\beta$.  Note that this
condition is only satisfiable if the quasienergy spectrum has no
degeneracies.  Then, it is possible to replace the $\T$-periodic
coefficients in Eq.~\eqref{3.15a} by their time-averages. Furthermore,
we may assume that in the long-time limit, the solution becomes
diagonal and time-independent, i.e., we make the ansatz
\begin{equation}
  \label{RWAansatz}
  P_{\alpha\beta}(t)=\mathrm{const.} = \delta_{\alpha\beta}\, f_\alpha\ .
\end{equation}
With these approximations, the quantum kinetic equation~\eqref{3.15a}
assumes the form
\begin{equation}
  \label{RWAkinetic}
  \begin{split}
  0 =
  \dot{f}_\alpha = &
  -
  w^\leads_{\alpha}
  f_\alpha
  +
  s^\leads
  \\
  &
  +
  \sum_{\alpha'}
  w^\ph_{\alpha\alpha'}\,(1-f_\alpha)\, f_{\alpha'}
  -\sum_{\alpha'}
  w^\ph_{\alpha'\alpha}\,(1-f_{\alpha'})\, f_{\alpha}\ .
  \end{split}
\end{equation}
Here, the population of the Floquet states~$\alpha$ due to the
wire-lead coupling is determined by the decay rates
\begin{equation}
  \label{3.42}
  w^\leads_{\alpha} \ldef
  \sum_{\ell=1}^L \sum_k
  \left|\langle\ell|\Phi_{\alpha,k}\rangle\right|^2\,
  \Gamma_\ell(\epsilon_{\alpha,k})
\end{equation}
and the source terms
\begin{equation}
  \label{3.47}
  s^\leads_\alpha \ldef
  \sum_{\ell=1}^L
  \sum_{k}
  \left|\langle\ell|\Phi_{\alpha,k}\rangle\right|^2\,
  \Gamma_\ell(\epsilon_{\alpha,k}) \, f(\epsilon_{\alpha,k}-\mu_\ell) .
\end{equation}
In addition, the electron-vibrational interaction contributes a
quantum Boltzmann type collision term to Eq.~\eqref{RWAkinetic}, which
takes into account the Pauli principle by the blocking factors
$1-f_\alpha$.  The corresponding scattering rates from one
state~$\alpha'$ into another state $\alpha$ are given by
\begin{equation}
  \label{3.47a}
  w^{\ph}_{\alpha \alpha'}
  \ldef
  2
  \sum_\nu
  \sum_{k}
  |\bar X^\nu_{\alpha\alpha',k}|^2\,
  N_\nu(\epsilon_{\alpha}-\epsilon_{\alpha'} + k\hbar\Omega)
  \ .
\end{equation}
%
\subsection{Thermal equilibrium}
\label{sec:3:thermaleq}
%
An important consistency check of the present theory is a thermal
equilibrium situation, where $H_{n n'}$ is time-independent and
where no external bias is present ($\mu_\ell=\mu$ for all $\ell$)
\cite{on_temperature}.
One can show that the diagonal ansatz~\eqref{RWAansatz}
leads to an \textit{exact} stationary solution of the kinetic
equation~\eqref{3.15a}. The populations $f_\alpha$ obey also the
kinetic equation~\eqref{RWAkinetic}, but with the Floquet states and
quasienergies replaced by the (adiabatic wire) eigenstates
$|\varphi_\alpha \rangle$ and the eigenenergies $E_\alpha$,
respectively, of the static Hamiltonian~$H_{n n'}$.  Moreover,
only $k=0$ contributes to the rates~\eqref{3.42}--\eqref{3.47a}.

Thermal equilibrium with respect to the coupling to the vibrational subsystems
is characterized by the detailed balance condition
\begin{equation}
  \label{detailedbalance}
  w^{\ph}_{\alpha\alpha'}\,(1-f_\alpha)\, f_{\alpha'}
  =
  w^{\ph}_{\alpha'\alpha}\,(1-f_{\alpha'})\, f_{\alpha}\ ,
\end{equation}
where the rates $w^{\ph}_{\alpha\alpha'}$ satisfy
\begin{equation}
  \label{3.49}
  \frac{w^{\ph}_{\alpha\alpha'}}{w^{\ph}_{\alpha'\alpha}} =
  \e^{(E_{\alpha'}-E_\alpha)/k_\mathrm{B} T}\ .
\end{equation}
A solution of Eqs.~\eqref{detailedbalance} and \eqref{3.49} is given
by the Fermi distribution $f_\alpha=f(E_\alpha-\mu')$, where the
chemical potential~$\mu'$ remains undetermined. It is due to the
additional condition $w^\leads_{\alpha} f_\alpha = s^\leads$ that
$\mu'$ adjusts to the chemical potential $\mu$ of the leads.  Thus,
the equilibrium solution of the kinetic equations~\eqref{3.15a} reads
\begin{equation}
  \label{3.49a}
  f_\alpha=f(E_\alpha-\mu)\,,
\end{equation}
in accordance with elementary principles of statistical physics of
non-interacting fermions.

\subsection{Numerical solution of the kinetic equation}

\label{sec:3:num}

For the solution of the nonlinear kinetic equation~\eqref{3.15a} one generally
has to resort to numerical methods.  We therefore use a propagation scheme for
the computation of the long-time limit of the solutions of the set of
non-linear equations~\eqref{3.15a}, which however, is numerically
rather time-consuming and, especially in the strongly driven regime,
only applicable for not too large systems.  We then verify the
$\T$-periodicity of the resulting solution and compute its Fourier
decomposition
\begin{equation}
  \label{3.50}
  \erho_{\alpha\beta}(t) =
  \sum_k \e^{-\i k\Omega t}
  \erho_{\alpha\beta,k}\,.
\end{equation}
As we will see in the subsequent section, the Fourier coefficients
\begin{equation}
\erho_{\alpha\beta,k} \ldef
  \frac{1}{\T}
  \int_0^\T
  \!\d t\,
  \e^{\i k\Omega t}\,
  \erho_{\alpha\beta}(t)
\end{equation}
fully specify the stationary current through the wire.
%
\section{Current through the wire}
\label{sec:current}
%
The net (incoming minus outgoing) current through contact $\ell$ is given by the negative
time derivative of the electron number
$N_\ell=\sum_{\bf k} c^\dagger_{{\bf k}\ell} c_{{\bf k} \ell}$ in
lead~$\ell$ multiplied by the electron charge~$-e$,
\begin{equation}
\label{I_L_start}
I_\ell(t)  = e \frac{\d}{\d t} \langle N_\ell\rangle_t =
e \, \frac{\i}{\hbar} \,\big\langle [H(t), N_\ell]\big\rangle_t\ .
\end{equation}
For the Hamiltonian~\eqref{3.1}, the commutator in Eq.~\eqref{I_L_start} is
readily evaluated to read
\begin{equation}
\label{I_L}
I_\ell(t)  = -
\frac{2e}{\hbar} \mathop{\mathrm{Im}} \sum_{\bf k} V_{{\bf k} \ell}
\langle c^\dagger_{{\bf k}\ell} c_\ell \rangle_t\ .
\end{equation}
For a weak wire-lead coupling, one can assume for all times a
factorization of wire and lead degrees of freedom.  This assumption
allows one to derive from Eq.~\eqref{I_L} an explicit expression for
the stationary, time-dependent net electrical current through the
contact~$\ell$ in terms of one-particle expectation values of the wire
electrons,
\begin{equation}
\label{current_general}
\begin{split}
I_\ell(t)  =
\frac{e}{\pi\hbar}\mathop{\rm Re}\int\limits_0^\infty \!\d\tau \! \int
\!\d\epsilon\,
&
\,\Gamma_\ell(\epsilon) \,\e^{\i\epsilon\tau/\hbar}\,
\Big\{
\big\langle c_\ell^\dagger\, \tilde c_\ell(t,t-\tau)\big\rangle_{t-\tau}
\\ &
-[c^\dagger_\ell,\tilde c_\ell(t,t-\tau)]_+ f(\epsilon-\mu_\ell)
\Big\}\ .
\end{split}
\end{equation}
Note that the anti-commutator $[c^\dagger_\ell,\tilde
c_\ell(t,t-\tau)]_+$ is in fact a c-number, which by means of the
transformation~\eqref{3.13} and the interaction picture
dynamics~\eqref{3.13a} of the wire operators in the Floquet picture
reads
\begin{equation}
  \label{anticomm}
  [c^\dagger_{\ell}, \tilde c_\ell(t,t-\tau)]_+ =
  \sum_\alpha
    \e^{-\i\epsilon_\alpha\tau/\hbar}\,
    \langle \Phi_\alpha(t-\tau)|\ell\rangle
    \langle \ell|\Phi_\alpha(t)\rangle\ .
\end{equation}
Similarly, the expectation value appearing in the current formula~\eqref{current_general}
can be expressed in terms of the density-matrix elements~\eqref{3.15} as
\begin{equation}
  \label{expectationvalue}
  \begin{split}
  \big\langle c_\ell^\dagger\, \tilde c_\ell(t,t-\tau)\big\rangle_{t-\tau} =
  \sum_{\alpha\beta}
  \e^{-\i\epsilon_\alpha\tau/\hbar}\,&
  \langle\Phi_\beta(t-\tau)|\ell\rangle\langle\ell|\Phi_\alpha(t)\rangle
  \\&\times
  \erho_{\alpha\beta}(t-\tau)\,.
  \end{split}
\end{equation}
These relations together, together with the spectral decompositions of
the Floquet states and of the density matrix, Eqs.~\eqref{2.8b} and
\eqref{3.50}, respectively, allow one to carry out the time and energy
integrals in expression~\eqref{current_general}.  The current
$I_\ell(t)$ obeys the time-periodicity of the driving field.
However, because we consider here excitations with frequencies in the
optical or infrared spectral range, the only experimentally accessible
quantity is the time-averaged current.  In the wide-band
limit---the extension to the general case is straightforward---we thus
obtain
\begin{equation}
\label{dc_current_l}
\begin{split}
\bar{I}_\ell =
e\Gamma_\ell\sum_{\alpha k}\Big[&
\sum_{\beta k'}
 \langle \Phi_{\beta,k'+k} |\ell\rangle\langle\ell|\Phi_{\alpha,k'} \rangle
 \erho_{\alpha\beta,k}\\
 &
-
  |\langle\ell|\Phi_{\alpha,k}\rangle|^2
  f(\epsilon_{\alpha,k}-\mu_\ell)
\Big]\,.
\end{split}
\end{equation}

Charge conservation, of course, requires that the charge on the wire
$Q_{\mathrm{wire}}(t)$ can only change by a current flow, amounting to
the continuity equation $\dot Q_{\mathrm{wire}}(t)=\sum_{\ell=1}^L
I_\ell(t)$.  Since the charge on the wire is bounded, the long-time
average of $\dot Q_{\mathrm{wire}}(t)$ must vanish.  From the
continuity equation one then finds
\begin{equation}
  \label{currentcons}
  \sum_{\ell=1}^L\bar I_\ell=0\ .
\end{equation}
For a two-terminal configuration, $\ell=\mathrm{L},\mathrm{R}$, we can
then introduce the time-averaged current
\begin{equation}
\bar I \ldef  \bar I_\mathrm{L} = -\bar I_\mathrm{R}.
\label{barI}
\end{equation}

To close this section we consider the thermal equilibrium situation
described in Sect.~\ref{sec:3:thermaleq}. It is characterized by
$\mu_\ell=\mu$ (absence of any transport voltage) and
$|\Phi_{\alpha,k}\rangle=0$ unless $k=0$ (absence of driving).
Inserting the equilibrium solution $P_{\alpha\beta,k} =
\delta_{\alpha\beta} \,\delta_{k,0}\, f(E_\alpha-\mu)$ into
Eq.~\eqref{dc_current_l}, we immediately see that all currents $\bar
I_\ell$ vanish---despite any possible asymmetry of the molecule itself
or of its coupling to the environment. Thus, the results of our
microscopic theory are in accordance with the second law of
thermodynamics.

%
\section{Control of currents in two-site systems}
\label{sec:applications}
%
With the necessary formalism at hand, we study in this section two
aspects of driven transport for which dissipation plays a significant
role: (i) the generation of pump currents by means of harmonic mixing
fields and (ii) optical current switching.  As a rather generic model,
which still captures the essential physics, we employ a symmetric wire
that consists of two sites in the two-terminal configuration sketched
in Fig.~\ref{fig:wiremodel_pump_mixing}.  The wire sites are coupled
by a hopping matrix element~$\Delta$ while an electromagnetic field
causes time-dependent shifts of the on-site energies.  Then, the wire
Hamiltonian reads
\begin{equation}
  \label{eq:twositerect}
  H_\wire(t)
  = -\Delta(c_\mathrm{L}^\dagger c_\mathrm{R}+c_\mathrm{R}^\dagger c_\mathrm{L})
  + \frac{a(t)}{2}(c_\mathrm{L}^\dagger c_\mathrm{L}-c_\mathrm{R}^\dagger c_\mathrm{R})
  \,,
\end{equation}
where $a(t)=a(t+\T)$ represents the dipole force on the wire electron
multiplied by the distance of the two wire sites.
Furthermore, we assume that the molecule couples equally strong to both
leads, thus, $\Gamma_\mathrm{L}=\Gamma_\mathrm{R}=\Gamma$.

In a realistic wire molecule, $\Delta$ is of the order $0.1\,{\rm
  eV}$. Thus, a wire-lead coupling strength $\Gamma=0.1 \Delta/\hbar$
corresponds to a current $e\Gamma=2.56\times10^{-5}$\,Amp\`ere and a
laser frequency $\Omega=\Delta/\hbar$ lies in the infrared spectral
range.  Furthermore, for a distance of {$1$\AA} between two
neighboring sites, a driving amplitude $A=\Delta$ is equivalent to an
electrical field strength of $10^7\,\mathrm{V/cm}$.

It turns out that for the description of the effects discussed below,
the off-diagonal elements of the single-particle density matrix
$\erho_{\alpha\beta}(t)$, $\alpha\ne\beta$, play an essential role.
Therefore, we have to go beyond the rotating-wave
ansatz~\eqref{RWAansatz} and consequently employ the propagation scheme for
the full nonlinear kinetic equation~\eqref{3.15a} described in
Sect.~\ref{sec:3:num}.

\begin{figure}[tp]
  \centerline{\includegraphics[width=0.9\linewidth]{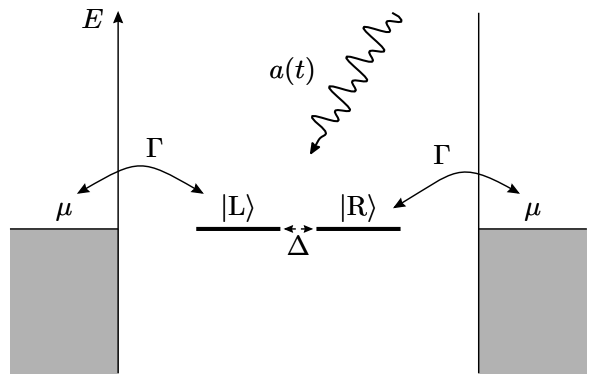}}
  \caption{
  Symmetric two-site structure coupled to two leads which rectifies an externally applied
  laser field of the form $a(t)=A_1\sin(\Omega t)+A_2\sin(2\Omega t + \varphi)$.  }
\label{fig:wiremodel_pump_mixing}
\end{figure}%

\subsection{Non-adiabatic pump current from harmonic mixing}
\label{subsec:mixing}

The Hamiltonian \eqref{eq:twositerect} with a driving of the form
$a(t)\propto\sin(\Omega t)$ has an intriguing symmetry, the so-called
generalized parity, which has been discussed widely in the context of
driven tunneling \cite{Grossmann1991a,Grossmann1991b}: a time translation
by half a driving period results for the external field in a minus
sign, i.e., $a(t+\T/2)=-a(t)$.  Thus, for the dipole coupling given in
the Hamiltonian \eqref{eq:twositerect}, the time shift $t\to t+\T/2$
is equivalent to interchanging the left and the right wire site.  In
addition, the dc current $\bar I$ is also inverted.  Consequently,
because $\bar I$ has to be independent of any (finite) time
translation, it must vanish \cite{Lehmann2003b}.  However, for a time-dependent
driving field of the form
\begin{equation}
\label{mixing}
a(t) = A_1\sin(\Omega t) + A_2\sin(2\Omega t+\varphi) ,
\end{equation}
with $A_1,A_2\neq 0$, the generalized parity is no longer present and
a non-adiabatic pump current emerges from the harmonic mixing of the
two driving frequencies \cite{Lehmann2003b}.  Its magnitude is
generally proportional to the coupling strength between wire and lead,
$\bar I\propto\Gamma$, with a prefactor that depends on the details
like the phase lag $\varphi$ or the amplitudes $A_1,A_2$.
The phase lag $\varphi=0$ represents a particular case because for
this value the wire Hamiltonian obeys time-reversal parity, i.e., it
is invariant under the operation
$(\mathrm{L},\mathrm{R},t)\to(\mathrm{R},\mathrm{L},-t)$.  As a
consequence, one finds that the dc current vanishes to linear order in
$\Gamma$ such that $\bar I\propto\Gamma^2$ \cite{Lehmann2003b}.
Figure \ref{fig:pump_mixing_gamma} demonstrates this behavior and,
moreover, reveals that already small phase lags of the order
$\varphi\approx0.001$ are sufficient to alter the qualitative
$\Gamma$-dependence of the dc current.
\begin{figure}[tp]
  \centerline{\includegraphics[width=0.9\linewidth]{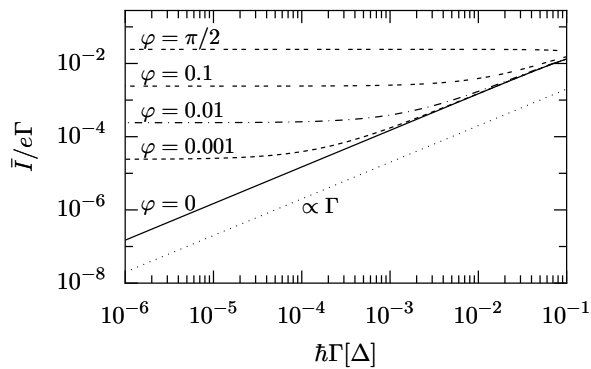}}
  \caption{
  Average current through the two-site wire from
  Fig.~\ref{fig:wiremodel_pump_mixing} driven by the harmonic mixing
  signal~\eqref{mixing} with amplitudes $A_1=2A_2=\Delta$ as a
  function of the wire-lead coupling strength $\Gamma$ (for
  $\kappa=0$) for different
  values of the phase difference $\varphi$. The
  driving frequency is $\Omega=\Delta/\hbar$,
  and the temperature is $k_\mathrm{B}T=0.25\,\Delta$.  The dotted
  line is proportional to $\Gamma$, corresponding to a current that is
  proportional to $\Gamma^2$.}
\label{fig:pump_mixing_gamma}
\end{figure}%

Harmonic mixing has also been studied recently for the motion of a
quantum particle in an infinitely extended tight-binding lattice, both
in the purely coherent regime~\cite{Goychuk2000a,Goychuk2001a} and for incoherent,
sequential quantum between adjacent sites~\cite{Goychuk1998b,Goychuk2001a}.  It
turns out that the dependence of the current on the phase lag $\varphi$
is qualitatively different in these two limiting cases. This raises
the question how the phase dependence of the current changes as
a function of the dissipation strength.

Generally, quantum dissipation results from a coupling of the quantum
system to an environment---here, the metallic leads and the
vibrational modes. Figure~\ref{fig:pump_mixing_phi}a depicts the
influence of only the wire-lead coupling: For a very weak coupling
strength $\Gamma=0.001\Delta/\hbar$ and $\kappa=0$ we find a dc
current proportional to $\sin\varphi$. With increasing $\Gamma$, the
dependence on $\varphi$ shifts towards $\cos\varphi$. In order to
investigate the influence of the vibrational coupling, we choose again
$\Gamma=0.001\Delta/\hbar$ and a finite but small vibrational coupling
strength~$\kappa$. The $\varphi$-dependence of the pump current is
given in Fig.~\ref{fig:pump_mixing_phi}b; it exhibits the same
dissipation-induced shift.
\begin{figure}[tp]
  \centerline{\includegraphics[width=0.9\linewidth]{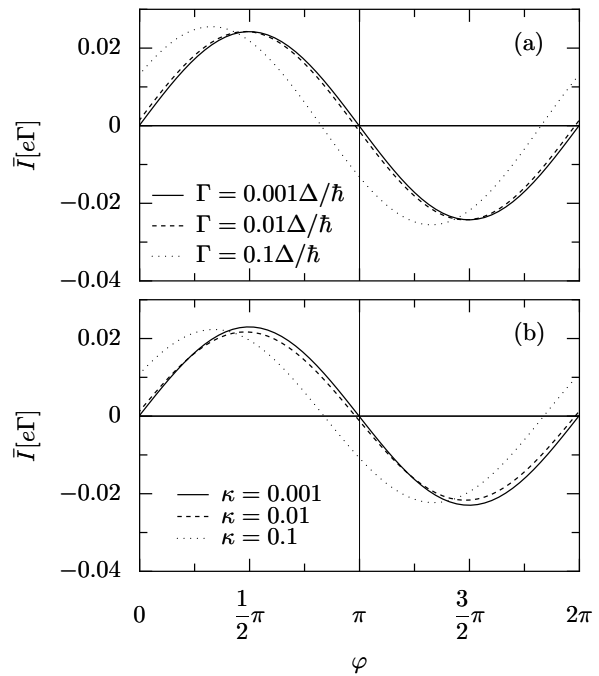}}
  \caption{Average current through the two-site wire sketched in Fig.~\ref{fig:wiremodel_pump_mixing}
  driven by the harmonic mixing signal~\eqref{mixing} as a function of
  the phase difference $\varphi$ (a) for different wire-lead coupling
  strengths $\Gamma$ (for $\kappa=0$) and (b) for different
  electron-vibrational coupling~$\kappa$ (for
  $\Gamma=0.001\Delta/\hbar$).  All other parameters are as in
  Fig.~\ref{fig:pump_mixing_gamma}.  }
\label{fig:pump_mixing_phi}
\end{figure}%

Interestingly enough, the electron-vibrational coupling can enhance
the pumping effect.  This enhancement is most pronounced in the
presence of the time-reversal parity discussed above, i.e., for
$\varphi=0$.  Figure~\ref{fig:pump_mixing_kappa} shows the pump
current as a function of the vibrational coupling strength $\kappa$.
We find that the dc current can be increased by more than one order of
magnitude. For values $\kappa\ll\hbar\Gamma/\Delta$, the
main dissipation comes from the leads and the vibrations are
practically without influence. Correspondingly, one is back to a the
situation of Fig.~\ref{fig:pump_mixing_gamma}, where the pump current
is proportional to $\Gamma^2$.
\begin{figure}[tp]
  \centerline{\includegraphics[width=0.9\linewidth]{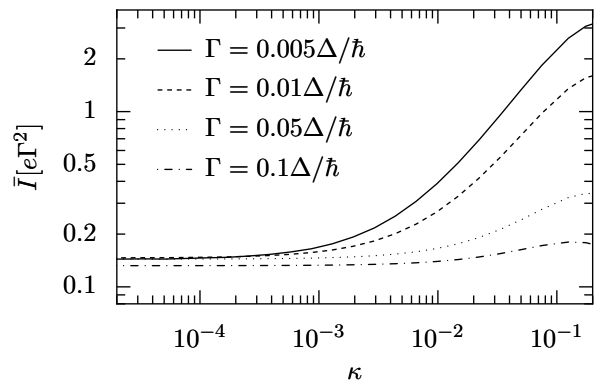}}
  \caption{Average current (in units of $e\Gamma^2$) through the two-site
  wire sketched in Fig.~\ref{fig:wiremodel_pump_mixing} driven by the
  harmonic mixing signal~\eqref{mixing} with amplitudes
  $A_1=2A_2=\Delta$ and phase difference $\varphi=0$ as a function
  of the electron-vibrational coupling strength~$\kappa$. Different
  values of the wire-lead coupling strength $\Gamma$ are shown.
}
\label{fig:pump_mixing_kappa}
\end{figure}%

\subsection{Laser-switched current gate}
\label{subsec:current_gate}

An external driving field can not only induce a pump current through
the molecular wire, but for proper parameters can also cause the
opposite effect: A driving of the shape $a(t)=A\sin(\Omega t)$ can
suppress almost completely the dc current even in the presence of a
large transport voltage $V$ \cite{Lehmann2003a}.  The physics behind
these suppressions is the so-called coherent destruction of tunneling
(CDT) that has been found in the context of tunneling in
time-dependent bistable potentials \cite{Grossmann1991a,Grossmann1991b,
Grossmann1992a, Goychuk1996a, Grifoni1998a}.  The central phenomenon
observed there is that for a driving with amplitude and frequency such
that the ratio $A/\hbar\Omega$ equals a zero of the Bessel
function $J_0$ (i.e., for the values 2.405.., 5.520.., 8.654.., \ldots), the
coherent tunneling dynamics comes to a standstill
\cite{Grossmann1992a}.  As a related effect for the transport through
such a tunnel system, one finds pronounced suppressions of the dc
current \cite{Lehmann2003a}.  The fact that coherent destruction of
tunneling is disrupted by finite dissipation \cite{Dittrich1993a,
  Morillo1993a, Kayanuma1993a, Grifoni1995a, Grifoni1998a, Fonseca2004a}, motivates our
investigation of the influence of dissipation on these current
suppressions.

We model the transport voltage $V$ by shifting the chemical potential
of the left (right) lead, $\mu_\mathrm{L}$ ($\mu_\mathrm{R}$), by
$-eV/2$ ($+eV/2$), cf.\ Fig.~\ref{fig:wiremodel_gate}. Due to the external voltage,
the electric field can in principle also cause a
static bias to the wire levels.  We do not take this
effect into account in the present work, but remark only that in the absence of a
vibrational coupling, the current suppressions are stable against an
internal bias~\cite{Lehmann2003b}.
\begin{figure}[tp]
  \centerline{\includegraphics[width=0.9\linewidth]{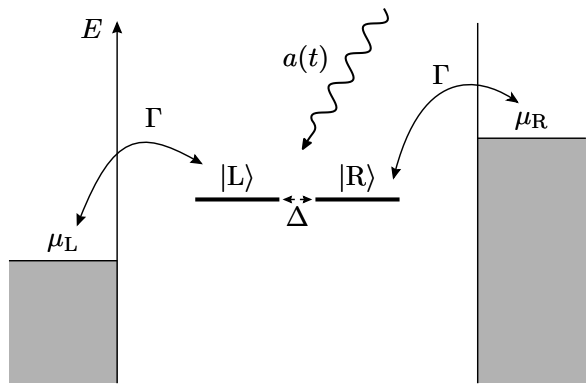}}
  \caption{Two-site wire attached to leads with the chemical potential difference
$\mu_\mathrm{R}-\mu_\mathrm{L}=eV$.
\label{fig:wiremodel_gate}}
\end{figure}%

Before focussing on the influence of electron-vibrational coupling, let us
first substantiate the discussion of the current suppressions by the
numerical results depicted in Fig.~\ref{fig:gate_current_unbiased}.  The
time-averaged current~$\bar I$ as a function of the laser amplitude~$A$
exhibits, besides a global decay, pronounced minima whenever the CDT
condition is fulfilled, i.e., when the ratio $A/\hbar\Omega$ assumes a zero
of the Bessel function $J_0$.  However, the current does not vanish
exactly, but a residual current remains; its value is proportional to the
molecule-lead coupling~$\Gamma$, as can be appreciated from the inset of
Fig.~\ref{fig:gate_current_unbiased}.  Since the current in the undriven
situation is also proportional to~$\Gamma$, we thus can conclude that the
maximal suppression is determined by a factor which is \textit{independent} of
$\Gamma$.

The inspection of the quasienergy spectrum reveals that CDT is related to
crossings of the quasienergies \cite{Grossmann1992a}.  Thus, at the center
of the current suppressions, the quasienergies are degenerate and the
condition \eqref{3.34} for the applicability of the rotating-wave
approximation is \textit{not} fulfilled.  Indeed, the dashed line in
Fig.~\ref{fig:gate_current_unbiased} demonstrates that a rotating-wave
approximation fails completely in the vicinity of current suppressions.
\begin{figure}[t]
  \centerline{\includegraphics[width=0.9\linewidth]{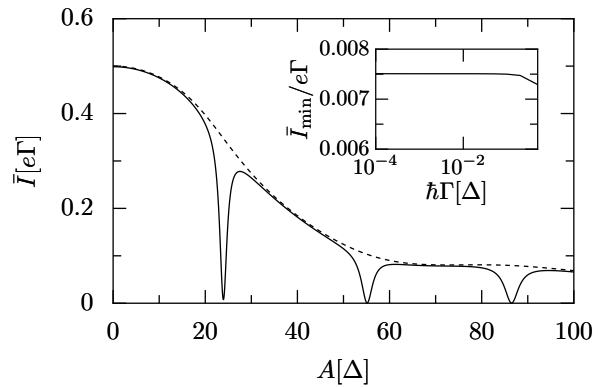}}
  \caption{Average current (solid) vs.\ driving amplitude for the setup
    sketched in Fig.~\ref{fig:wiremodel_gate}.  The leads'
    chemical potentials are $\mu_\mathrm{R}=-\mu_\mathrm{L}=10 \Delta$; the
    other parameters read $\hbar\Omega=10\Delta$, $k_\mathrm{B} T=0.25\Delta$,
    $\hbar\Gamma=0.1\Delta$.  The dashed line marks the result obtained within
    a rotating-wave approximation.  The inset depicts the minimal current $\bar
    I_\mathrm{min}$ at the first suppression as a function of the wire-lead
    coupling strength~$\Gamma$.}
  \label{fig:gate_current_unbiased}
\end{figure}%

A central question to be addressed is the robustness of the current
suppressions against dissipation.  In the corresponding tunneling problem, the
CDT driving alters both the coherent and the dissipative time scale by the
same factor \cite{Fonseca2004a}.  Thus, one might speculate that a
vibrational coupling leaves the effect of the driving on the current
qualitatively unchanged.  Figure~\ref{fig:gate_current_biased_dissipation},
however, demonstrates that this is not the case.  With increasing
dissipation strength $\kappa$, the characteristic current suppressions become washed out
until they finally disappear when $\kappa$ becomes of the order unity.  This
detracting influence underlines the importance of quantum coherence for the
observation of those current suppressions.  Note that the dissipation affects
only the depth of the suppressions while the width remains unchanged.
We close this section with the remark that within the present setup of
two driven tunnel-coupled orbitals (cf.\ 
Figure~\ref{fig:wiremodel_gate}) and within our chosen parameter range,
we do not detect the analogue of the effect of a stabilization of CDT
within a certain temperature range
\cite{Dittrich1993a,Dittrich1993b,Oelschlaegel1993a,Thorwart2000a,Thorwart2001a,Makarov1995a}
or, likewise, with increasing external noise \cite{Grossmann1993a}, as it has
been reported for driven, dissipative symmetric bistable systems.
\begin{figure}[tp]
  \centerline{\includegraphics[width=0.9\linewidth]{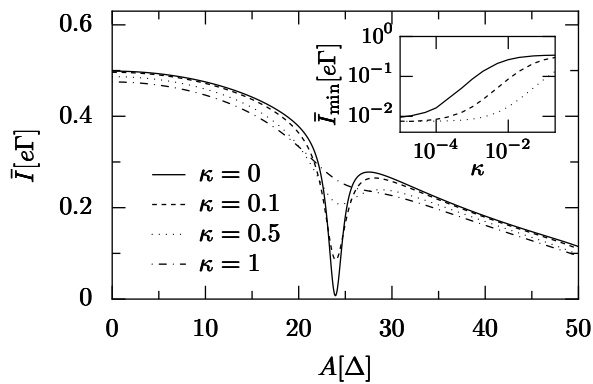}}
  \caption{Average current vs.\ driving amplitude for the setup sketched
    in Fig.~\ref{fig:wiremodel_gate} in the presence of dissipation of
    the form~\eqref{3.2l}.  The inset depicts the minimal current
    $\bar I_\mathrm{min}$ at the first suppression as a function of
    the electron-phonon coupling strength~$\kappa$ for
    $\Gamma=0.001\Delta/\hbar$ (solid line), $0.01\Delta/\hbar$
    (dashed), and $0.1\Delta/\hbar$ (dotted).  All other parameters
    are as in Fig.~\ref{fig:gate_current_unbiased}.}
  \label{fig:gate_current_biased_dissipation}
\end{figure}%

%
%
\section{Conclusions}
%
%
We have derived a nonlinear quantum kinetic equation that allows one
to investigate for a molecule the simultaneous influence of a laser
field, a coupling to leads, and in addition, a coupling to
vibrational modes.  The use of Floquet states as a basis set for the
reduced single-particle density matrix represents a most important
technical cornerstone.  It enables one both the exact inclusion
of the driving field and an efficient treatment of the dissipative
couplings.  Since the vibrational modes provide an \textit{effective
electron-electron interaction}, a formalism for general situations
requires one to resort to further approximations such as a
Hartree-Fock decoupling scheme.

Within this kinetic equation formalism, we have investigated the
influence of quantum dissipation on recently proposed transport
effects caused by the action of laser fields on molecular wires.
For the non-adiabatic electron pumping that emerges from harmonic
mixing, we find that dissipation can play a constructive role to the
extent that it can significantly enhance the current.

For the model under investigation, we observed an enhancement of
the pump current by more than one order of magnitude.  Moreover, the
present scheme allows one to trace back the dependence of the pump
current on the phase lag between the two harmonic mixing fields to the
increasing influence of dissipation.
The situation is less promising for effects that depend intrinsically
on quantum coherence.  We have found that the coherent current
suppressions are derogated by the coupling to vibrational modes.
Nevertheless, the effect persists provided that the quantum dynamics
of the wire electrons remains predominantly coherent.
Finally, we share the hope that our general theoretical findings will provide
motivation and prove useful to experimentalists to initiate corresponding,
tailored experiments on driven molecular wires in the not too distant future.

\section{Acknowledgement}

This work has been supported by the Volkswagen-Stiftung under Grant
No.\ I/77 217 and the Deutsche Forschungsgemeinschaft through SFB 486,
project A10.



\begin{thebibliography}{10}
\bibitem{on_temperature}
Of course, thermal equilibrium requires all baths to be at a common
  temperature, as was put into our model from the very beginning.
\end{thebibliography}
\end{document}